\documentclass[a4paper,12pt,english,oneside]{article}

\usepackage{graphicx}
\usepackage{amssymb,amsmath,scalefnt,cite}

\begin{document}

{\par\centering \bf MICROSCOPIC DERIVATION OF THE  ONE-QUBIT KRAUS OPERATORS FOR AMPLITUDE AND PHASE DAMPING \par}

\vspace{12mm}

{\par\centering \bf M. Arsenijevi\' c$^{1}$ and  N. Bankovi\' c $^{2}$\par}

\bigskip

\vspace{12mm}
\textit{$^{1}${Department of Physics, Faculty of Science, Kragujevac,
Serbia}}
{\par\centering e-mail: momirarsenijevic@kg.ac.rs\par}

\textit{$^{2}${Technical College of Applied Studies in Kragujevac,
Kragujevac, Serbia}}

\noindent{\bf ABSTRACT} This article presents microscopic derivation of the Kraus operators for (the generalized) amplitude and phase damping process.
Derivation is based on the recently developed method [Andersson et al, J. Mod.Opt. 54, 1695 (2007)] which concerns finite dimensional systems (e.g. qubit).  The form of  these operators is usually estimated without insight into the microscopic details of the dynamics. The behaviour of the qubit dynamics is simulated and depicted via Bloch sphere change.

\vspace{15mm}

{\centerline{\bf 1. INTRODUCTION}}
\vspace{12mm}

Quantum information processing substantially depends on the
mathematical details of the environmental influence exerted on the
qubit-registers \cite{N&Ch}. Nevertheless, to the best of our
knowledge, theoretical origin of the widely used Kraus operators
for the one-qubit quantum noise-channels  \cite{N&Ch,RomLoFranco,BCSQIQC,QCD,ByChManiscalco,Marinescu,KlimSoto} has not been investigated yet. In this paper we perform a thorough  analysis of the
microscopic models for the standard one-qubit amplitude damping
and phase damping quantum processes. To this end we use a recently
formulated method \cite{AndCresHall} for derivation of the Kraus operators
from a microscopic master-equation description of the processes.
We find the unitary equivalence of the here derived Kraus
operators with those widely used in the literature, thus presenting
the same quantum noise process. The converse conclusion regarding
the one-qubit depolarizing process will be presented elsewhere
\cite{GDP}.

In Section  2, we overview the method of Andersson et al \cite{AndCresHall}. In Section 3
we derive the Kraus operators for the generalized amplitude damping process,
while in Section 4 we derive the Kraus operators for the phase damping
process. Section 5 is the  conclusion.

\vspace{15mm}

{\centerline{\bf 2. OVERVIEW OF THE  METHOD}}
\vspace{12mm}

In the paper \cite{AndCresHall}, the authors  developed a general
procedure for deriving a Kraus decomposition from  the known master
equation and vice versa, regarding the  finite-dimensional quantum
systems. The only assumption is that the master equation is local
in time.

The so-called Nakajima-Zwanzig projection method
\cite{BreuPetr,RivasHuelga} gives the following master equation
for the system's density operator $\hat{\rho}_S(t)$, ($\hbar=1$):
\begin{equation}\label{NZmaster}
\frac{d\hat{\rho}_S(t)}{dt}=-i[\hat{H},\hat{\rho}_S(t)]+\int_{0}^{t}
\mathcal{K}_{t,s}[\hat{\rho}_S(s)]ds,
\end{equation}
where $\hat{H}$ represents the system's self-Hamiltonian (that
includes the so-called Lamb-shift term) and $\mathcal{K}_{t,s}$ is
the memory kernel which accounts for the non-unitary effects  due
to  the environment.

Certain processes can be written  in a local-in-time form
\cite{RivasHuelga,BreuPetr}:
\begin{equation}\label{Localmaster}
\dot{\hat{\rho}}_{S}(t)=\Lambda_{t}(\hat{\rho}_{S}(t)),
\end{equation}
where $\Lambda_{t}$ is a linear map which preserves hermiticity,
positivity and  unit trace of $\hat{\rho}_{S}(t)$ and has the
property:
\begin{equation}\label{Traceproperty}
tr\Lambda_{t}(\hat{\rho}_{S}(t))=0.
\end{equation}

Alternatively, dynamics can be presented in a non-differential,
``integral'' form \cite{BreuPetr,RivasHuelga,AndCresHall}:
\begin{equation}\label{Integralform}
\hat{\rho}_{S}(t)=\phi_{t}(\hat{\rho}_{S}(0)),
\end{equation}
where $\phi_{t}$ is a completely positive and trace preserving
linear map.

It can be shown \cite{AndCresHall} that linear maps $\Lambda_{t}$
and $\phi_{t}$ are connected via the matrix differential equation:
\begin{equation}\label{IntDiffconnection}
\dot{F}=LF,
\end{equation}
where the matrix elements of $L$ are given by:
\begin{equation}\label{MatrixelemL}
L_{kl}=tr[\mathcal{G}_k\Lambda(\mathcal{G}_l)].
\end{equation}
In eq.\eqref{MatrixelemL}, $\{\mathcal{G}_k\}$ is  any orthonormal
basis  of the Hermitian operators acting on the system's Hilbert
space. For the time independent $\Lambda_t$, i.e. $L$,
eq.\eqref{IntDiffconnection} has the unique solution:
\begin{equation}\label{Integralformsolution}
F=e^{Lt}.
\end{equation}

Complete positivity of the map $\phi_t$ (and hence of the matrix
$F$) is equivalent to the positivity of the, so called, Choi
matrix, $S$ \cite{Choi,AndCresHall}, whose elements are defined as
\cite{AndCresHall}:
\begin{equation}\label{Choimatrix}
S_{nm}=\sum_{s,r}F_{sr}tr[\mathcal{G}_r\mathcal{G}_n^{\dag}\mathcal{G}_s\mathcal{G}_m].
\end{equation}

With the use of equation \eqref{Choimatrix},
eq.\eqref{Integralform} takes the form:
\begin{equation}\label{AlternDecomp}
\phi(\hat{\rho}_S(0))=\sum_{nm}S_{nm}\mathcal{G}_n\hat{\rho}_S(0)\mathcal{G}_m^{\dag},
\end{equation}
which, after diagonalization of the  $S$ matrix:
\begin{equation}\label{ChoiDiag}
S=UDU^{\dag},
   \end{equation}
gives rise to a Kraus decomposition. The eigenvalues $d_i$ and the
eigenvectors of the $S$ matrix constitute the diagonal matrix $D$
and the unitary matrix $U=(u_{ij})$ respectively; columns of the
unitary $U$ operator are  the normalized eigenvectors of the $S$
matrix. Then the Kraus operators:
\begin{equation}\label{KrausoperViaDandU}
E_i=\sum_j\sqrt{d_i}u_{ji}\mathcal{G}_j
   \end{equation}
yield the Kraus decomposition of the  dynamical map $\phi_t$:
\begin{equation}\label{KrausDecompphiB}
\phi_t(\hat{\rho}_S(0))=\sum_{k}\hat{E}_k(t)\hat{\rho}_S(0)\hat{E}_k^{\dag}(t).
\end{equation}

Therefore, the chain of the construction is established: from a master
equation to calculate $L$, then via relation
\eqref{Integralformsolution} to obtain the matrix $F$ and, due to
eq.\eqref{Choimatrix} and diagonalization eq.\eqref{ChoiDiag} of
the Choi matrix to calculate the Kraus operators
eq.\eqref{KrausoperViaDandU}.

\vspace{15mm}

\centerline{\bf 3.  THE GENERALIZED AMPLITUDE DAMPING CHANNEL}

\vspace{12mm}

The standard master equation for the  amplitude damping process,
at absolute zero, $T=0$K, reads \cite{RomLoFranco}:
\begin{align}
\label{ADmaster}
 \frac{d\hat{\rho}_S(t)}{dt}&=\frac{\gamma}{2}(
   2\hat{\sigma}_{-}\hat{\rho}_S(t)\hat{\sigma}_{+}-\hat{\sigma}_{+}\hat{\sigma}_{-}
   \hat{\rho}_S(t)-\hat{\rho}_S(t)\hat{\sigma}_{+}\hat{\sigma}_{-})\,,
\end{align}
while the corresponding standard Kraus operators:
\begin{align}\label{ADkraus}
\hat{E}_{0}&=|0\rangle\langle0|+\sqrt{1-\lambda(t)}|1\rangle\langle1|,&
\hat{E}_{1}&=\sqrt{\lambda(t)}|0\rangle\langle1|
\end{align}
and the Pauli operators
$\hat{\sigma}_z=|0\rangle\langle0|-|1\rangle\langle1|$,
$\hat{\sigma}_x=|0\rangle\langle1|+|1\rangle\langle0|$,
$\hat{\sigma}_y=\imath|0\rangle\langle1|-\imath|1\rangle\langle0|$,
and
$\hat{\sigma}_{\pm}=\frac{1}{2}(\hat{\sigma}_x\pm\imath\hat{\sigma}_y)$.

To describe the amplitude damping process for all temperatures,
the following Hamiltonian is often regarded \cite{RivasHuelga}:
\begin{equation}\label{ADHamiltonian}
\hat{H}=\frac{\omega_0}{2}\hat{\sigma}_z+\int^{\omega_{max}}_0
d\omega \hat{a}^{\dag}_{\omega}\hat{a}_{\omega}+
\int^{\omega_{max}}_0 d\omega h(\omega)
(\hat{a}^{\dag}_{\omega}\hat{\sigma}_{-}+\hat{a}_{\omega}\hat{\sigma}_{+}).
\end{equation}

The first term on the right side of eq.\eqref{ADHamiltonian}
denotes the system's self-Ha\-mil\-to\-ni\-an, the second
denotes the self-Ha\-mil\-to\-ni\-an of the environment (a thermal
bath of linear non-interacting harmonic oscillators and
$\hat{a}_{\omega}$ representing the bosonic "annihilation"
operator for the frequency $\omega$) while the last term
represents the interaction with the coupling-coefficients
$h(\omega)$. $\omega_{max}$ is the 'cutoff frequency' for the
bath's oscillators; one may take the limit $\omega_{max}\to\infty$
providing that the $h(\omega)$ sufficiently quickly decreases.

As distinct from eq.\eqref{ADmaster}, the microscopic Markovian
master equation, in interaction picture, derived from  Hamiltonian
\eqref{ADHamiltonian} reads \cite{RivasHuelga}:
\begin{equation}\label{ADmastereq}
       \begin{array}{r c l}
\displaystyle\frac{d\hat{\rho}_S(t)}{dt}&=&
-i[(\displaystyle\frac{\Delta}{2}+\Delta')\hat{\sigma}_z,\hat{\rho}_S(t)]\\&
+&2\pi J(\omega_0)(\langle n(\omega_0)\rangle +
1)\Big[\hat{\sigma}_{-}\hat{\rho}_S(t)\hat{\sigma}_{+}-
\displaystyle\frac{1}{2}\{\hat{\sigma}_{+}\hat{\sigma}_{-},\hat{\rho}_S(t)\}\Big]\\&
+&2\pi J(\omega_0)\langle n(\omega_0)\rangle
\Big[\hat{\sigma}_{+}\hat{\rho}_S(t)\hat{\sigma}_{-}-
\displaystyle\frac{1}{2}\{\hat{\sigma}_{-}\hat{\sigma}_{+},\hat{\rho}_S(t)\}\Big]\,,
\end{array}
\end{equation}
where $\Delta=\mathrm{P.V.}\int_{0}^{\omega_\textrm{max}} d\omega'
\frac{J(\omega')}{\omega_0-\omega'}$  and
$\Delta'=\mathrm{P.V.}\int_{0}^{\omega_\textrm{max}} d\omega'
\frac{J(\omega')\langle n(\omega')\rangle}{\omega_0-\omega'}$
denote the Lamb-like shift and the Stark-like shift contributions
from the vacuum and the thermal field, respectively, and curly
brackets stand for anti-commutator. $\mathrm{P.V.}$ stands for the
Cauchy principal value of the integral. $J(\omega)$ represents the
spectral density of the bath.

The master equation \eqref{ADmastereq} reduces to the standard AD
master equation \eqref{ADmaster} for $T=0$K, and is therefore
often called generalized amplitude damping (GAD) channel. Below,
due to the procedure described in Section 2, from
eq.\eqref{ADmastereq} we derive the GAD Kraus operators, which
will turn out to be unitary equivalent with the known GAD Kraus
operators \cite{SriBan}:
\begin{eqnarray}
\label{GADKraus}
\begin{array}{ll}
E_0 \equiv \sqrt{p}\left[\begin{array}{ll} \sqrt{1-\lambda(t)} & 0
\\ 0 & 1
\end{array}\right]; ~~~~ &
E_1 \equiv \sqrt{p}\left[\begin{array}{ll} 0 & 0 \\
\sqrt{\lambda(t)} & 0
\end{array}\right];  \\
E_2 \equiv \sqrt{1-p}\left[\begin{array}{ll} 1 & 0 \\ 0 &
\sqrt{1-\lambda(t)}
\end{array}\right]; ~~~~ &
E_3 \equiv \sqrt{1-p}\left[\begin{array}{ll} 0 & \sqrt{\lambda(t)}
\\ 0 & 0
\end{array}\right],
\end{array}
\end{eqnarray}
where $\lambda(t) \equiv 1 - e^{-\gamma_0(2N_{\rm th} +1) t}; p
\equiv \frac{N_{\rm th}+1}{2N_{\rm th} +1}$. $N_{\rm
th}=\left[e^{(\omega/T)}-1\right]^{-1}$ while $\gamma$ appears in
\eqref{ADmaster}.

To ease the calculation, we introduce the following notation:
$\mathbf{x}=\displaystyle\frac{\Delta}{2}+\Delta'$,
$\mathbf{y}=2\pi J(\omega_0)(\langle n(\omega_0)\rangle + 1)>
\mathbf{z}=2\pi J(\omega_0)\langle n(\omega_0)\rangle\geq0$ with
which the equation \eqref{ADmastereq} reads:
\begin{equation}\label{ADmastereqxyz}
       \begin{array}{r c l}
\displaystyle\frac{d\hat{\rho}_S(t)}{dt}&=&-i
\mathbf{x}[\hat{\sigma}_z,\hat{\rho}_S(t)]\\&
+&\mathbf{y}[\hat{\sigma}_{-}\hat{\rho}_S(t)\hat{\sigma}_{+}-
\frac{1}{2}\{\hat{\sigma}_{+}\hat{\sigma}_{-},\hat{\rho}_S(t)\}]\\[0.5ex]&
+&\mathbf{z}[\hat{\sigma}_{+}\hat{\rho}_S(t)\hat{\sigma}_{-}-\frac{1}{2}
\{\hat{\sigma}_{-}\hat{\sigma}_{+},\hat{\rho}_S(t)\}]\,.
\end{array}
\end{equation}

Now, from \eqref{ADmastereqxyz} and using eq.(6) from the body
text, the $\hat{\sigma}_z$-representation  of the $L$ matrix takes
the form:
\begin{equation}\label{ADmatL}
L=\left(
\begin{array}{cccc}
 0 & 0 & 0 & 0 \\
 0 & \frac{1}{2} (-\mathbf{y}-\mathbf{z}) & -2 \mathbf{x} & 0 \\
 0 & 2 \mathbf{x} & \frac{1}{2} (-\mathbf{y}-\mathbf{z}) & 0 \\
 \mathbf{z}-\mathbf{y} & 0 & 0 & -\mathbf{y}-\mathbf{z}
\end{array}
\right)\,.
\end{equation}

In order to facilitate the calculation of the exponential $F$
matrix, we multiply the $L$ matrix by
$\frac{2}{(\mathbf{y}+\mathbf{z})}$ that allows introduction of
new variables: $\theta
=\frac{4\mathbf{x}}{\mathbf{y}+\mathbf{z}}$, $\Omega
=-\frac{2(\mathbf{y}-\mathbf{z})}{(\mathbf{y}+\mathbf{z})}$, $\tau
=\frac{(\mathbf{y}+\mathbf{z})}{2}t$; $\Omega\in[-2,0)$,
$\tau\in(-\infty,\infty)$. Then follows:
\begin{equation}\label{ADmatLthetaOmegatau}
\frac{2L}{\mathbf{y}+\mathbf{z}}=\left(
\begin{array}{cccc}
 0 & 0 & 0 & 0 \\
 0 & -1 & -\theta  & 0 \\
 0 & \theta  & -1 & 0 \\
 \Omega  & 0 & 0 & -2
\end{array}
\right)\,.
\end{equation}

and
\begin{equation}\label{ADmatF}
 F=e^{\frac{2L}{\mathbf{y}+\mathbf{z}}\tau},
\end{equation}
which obtains the form:
\begin{equation}\label{ADmatF1}
F=\left(
\begin{array}{cccc}
 1 & 0 & 0 & 0 \\
 0 & e^{-\tau } \cos (\theta  \tau ) & -e^{-\tau } \sin (\theta  \tau ) & 0 \\
 0 & e^{-\tau } \sin (\theta  \tau ) & e^{-\tau } \cos (\theta  \tau ) & 0 \\
 e^{-\tau } \Omega  \sinh (\tau ) & 0 & 0 & e^{-2 \tau }
\end{array}
\right)\,.
\end{equation}

From eq.\eqref{ADmatF1} we obtain the corresponding Choi matrix
(Section 2), whose diagonalization  gives the
following set of eigenvalues:
\begin{subequations}\label{ADEigenavalues}
\begin{align}
-\frac{1}{4} e^{-2 \tau} \left(-1+e^{2 \tau }\right) (-2+\Omega ), \\
\quad \frac{1}{4} e^{-2 \tau }
\left(-1+e^{2 \tau }\right) (2+\Omega ), \\[0.5ex]
\frac{1}{4} e^{-2 \tau } \left(2+2 e^{2 \tau }-\sqrt{16 e^{2 \tau
}+\Omega ^2-2 e^{2 \tau } \Omega ^2+e^{4 \tau } \Omega
^2}\right),\\ \frac{1}{4} e^{-2 \tau } \left(2+2 e^{2 \tau
}+\sqrt{16 e^{2 \tau }+\Omega ^2-2 e^{2 \tau } \Omega ^2+e^{4 \tau
} \Omega ^2}\right)\,,
\end{align}
\end{subequations}

\noindent and the  respective non-normalized eigenvectors:
\begin{subequations}\label{ADEigenvectors}
\center{
\begin{align}
 \left\{0,\frac{1}{2} i e^{-\tau } \left(-1+e^{2 \tau
}\right) \text{Csch}[\tau ],1,0\right\},\\\left\{0,-\frac{1}{2} i
e^{-\tau } \left(-1+e^{2 \tau }\right) \text{Csch}[\tau
],1,0\right\},\\ \left\{\frac{e^{-\tau } \left(-\sqrt{16 e^{2 \tau
}+\Omega ^2-2 e^{2 \tau } \Omega ^2+e^{4 \tau } \Omega ^2}+4
e^{\tau } \text{Cos}[\theta  \tau ]\right)}{2 (-2 i
\text{Sin}[\theta  \tau ]+\Omega  \text{Sinh}[\tau
])},0,0,1\right\},\\\left\{\frac{e^{-\tau } \left(\sqrt{16 e^{2
\tau }+\Omega ^2-2 e^{2 \tau } \Omega ^2+e^{4 \tau } \Omega ^2}+4
e^{\tau } \text{Cos}[\theta  \tau ]\right)}{2 (-2 i
\text{Sin}[\theta  \tau ]+\Omega \text{Sinh}[\tau
])},0,0,1\right\}\,.
\end{align}}
\end{subequations}

Hence we obtain the first two Kraus matrices for GAD, eq.(4):
\begin{equation}\label{AD1}
   \mathbb{E}_1=\left(
\begin{array}{cc}
 0 & 0 \\
 \frac{1}{2} i \sqrt{\left(e^{-2 \tau }-1\right) (\Omega -2)} & 0
\end{array}
\right),
\end{equation}

\begin{equation}\label{AD2}
 \mathbb{E}_2=\left(
\begin{array}{cc}
 0 & -\frac{1}{2} i \sqrt{\left(1-e^{-2 \tau }\right) (\Omega +2)} \\
 0 & 0
\end{array}
\right).
\end{equation}

By introducing:

\begin{equation}\label{Smena1}
   A=-2i\sin[\theta \tau ]+\Omega \sinh[\tau
],
\end{equation}

\begin{equation}\label{Smena2}
   B_{\pm }=e^{-4\tau }\left(2+2e^{2\tau }\pm \sqrt{\Omega
^2+e^{2\tau }\left(16+\left(-2+e^{2\tau }\right)\Omega
^2\right)}\right),
\end{equation}

\begin{equation}\label{Smena3}
    C_{\pm }=e^{-2\tau }\left(\sqrt{\Omega ^2+e^{2\tau
}\left(16+\left(-2+e^{2\tau }\right)\Omega ^2\right)}\pm 4e^{\tau
}\cos[\theta \tau ]\right)^2,
\end{equation}

\begin{equation}\label{Smena4}
    D=4e^{\tau -i\theta \tau }
\end{equation}
\noindent and
\begin{equation}\label{Smena5}
   E_{\pm }=\left(1-e^{2\tau }\right)\Omega \pm \sqrt{\Omega
^2+e^{2\tau }\left(16+\left(-2+e^{2\tau }\right)\Omega ^2\right)},
\end{equation}

\noindent another pair of  Kraus  matrices, $ \mathbb{E}_3$ and $
\mathbb{E}_4$, can be written as:
\begin{equation}\label{AD3}
 \mathbb{E}_3=\frac{\sqrt{\frac{|A|^2B_-}{4 |A|^2+C_-}}}{2 \sqrt{2} A}\left(
\begin{array}{cc}
 D-E_+ & 0 \\
 0 & D^*+E_-
\end{array}
\right)
\end{equation}
and
\begin{equation}\label{AD4}
 \mathbb{E}_4=\frac{\sqrt{\frac{|A|^2 B_+ }{4 |A|^2+C_+}}}{2 \sqrt{2}
A}\left(
\begin{array}{cc}
 D-E_- & 0 \\
 0 & D^*+E_+
\end{array}
\right)\,.
\end{equation}
It is straightforward yet tedious task to confirm the completeness
relation $ \sum_{k}\hat{E}_k(t)^{\dag}\hat{E}_k(t)=\hat{I}$
($\hat{I}$ is the identity operator) for the  Kraus matrices
eqs.\eqref{AD1}, \eqref{AD2}, \eqref{AD3} and \eqref{AD4}.

For the bath on $T=0K$, the parameters $\mathbf{x}=0=\mathbf{y}$,
equivalently $\theta = 0, \Omega = - 2$, the GAD master equation
eq.\eqref{ADmastereq} reduces to the standard AD master equation
eq.\eqref{ADmaster}. Now placing $\theta=0, \Omega=-2$ in
eqs.\eqref{AD1}-\eqref{AD4}, we obtain:
\begin{equation}\label{AD1Stand}
 \mathbb{E}_1=\left(
\begin{array}{cc}
 0 & 0 \\
 i \sqrt{1-e^{-2 \tau }} & 0
\end{array}
\right)\,,
\end{equation}
\begin{equation}\label{AD2Stand}
    \mathbb{E}_2=0,
\end{equation}

\begin{equation}\label{AD3Stand}
    \mathbb{E}_3=0,
\end{equation}
and
\begin{equation}\label{AD4Stand}
 \mathbb{E}_4=\left(
\begin{array}{cc}
 -e^{-\tau } & 0 \\
 0 & -1
\end{array}
\right)\,.
\end{equation}
It is easy to prove the completeness relation for the Kraus
matrices eqs.\eqref{AD1Stand} and \eqref{AD4Stand}.

The matrices $\mathbb{E}_1'=-i \mathbb{E}_1$ and
$\mathbb{E}_4'=-\mathbb{E}_4$  are unitary-equivalent with the
standard AD Kraus operators eq.\eqref{ADkraus}. That is, the sets
eqs.\eqref{AD1Stand}-\eqref{AD4Stand} and eq.\eqref{ADkraus}
describe the same process.

Unitary equivalence of the Kraus matrices \eqref{AD1}-\eqref{AD4}
and the  GAD Kraus matrices \eqref{GADKraus}  follows from the
following observations. First, the GAD Kraus matrices eq.(5)
reduce to  standard ones eq.\eqref{ADkraus} for $N_{\rm th}=0$
i.e. $p=1$ (in our notation these are: $\theta = 0, \Omega = -
2$). Second, for both sets of the Kraus operators,
\eqref{AD1}-\eqref{AD4}, and eq.\eqref{GADKraus}, it easily
follows:
\begin{subequations}
\label{ADEffects}
\begin{align}
\phi_{\tau}(\hat{I})&=\hat{I}+\frac{\Omega}{2}(1-e^{-2\tau})\hat{\sigma}_z\,, \label{ADEffectsA} \\
\phi_{\tau}(\hat{\sigma}_x)&=e^{-\tau}(\hat{\sigma}_x\cos\theta\tau+\hat{\sigma}_y\sin\theta\tau)\,, \label{ADEffectsB}\\
\phi_{\tau}(\hat{\sigma}_y)&=e^{-\tau}(\hat{\sigma}_y\cos\theta\tau-\hat{\sigma}_x\sin\theta\tau)\,, \label{ADEffectsC}\\
\phi_{\tau}(\hat{\sigma}_z)&=e^{-2\tau}\hat{\sigma}_z\,,
\label{ADEffectsD}
\end{align}
\end{subequations}
which, bearing in mind $\hat{\rho}=\frac{1}{2}\left(\hat{I}
+\vec{n} \cdot \hat{\vec{\sigma}}\right)$, gives rise to:
\begin{equation}\label{ADDensOperBlochthetaOmegatau}
\begin{array}{r c l}
  \phi_{\tau}(\hat{\rho})&=&\frac{1}{2}[\hat{I}+e^{-\tau}\sin v\cos(u+\theta\tau)\hat{\sigma}_x
  +e^{-\tau}\sin v\sin(u+\theta\tau)\hat{\sigma}_y\\&+&(\frac{\Omega}{2}(1-e^{-2\tau})+e^{-2\tau}\cos
  v)\hat{\sigma}_z]\,,
  \end{array}
\end{equation}
i.e. to:
\begin{equation}\label{ADDensOperBlochxyz}
\begin{array}{r c l}
 \phi_{t}(\hat{\rho})&=&\frac{1}{2}[\hat{I}+
 e^{-\frac{1}{2} t (\mathbf{y}+\mathbf{z})} \cos(u+2t\mathbf{x})\sin v
\hat{\sigma}_x + e^{-\frac{1}{2}t(\mathbf{y}+\mathbf{z})}\sin v
\sin(u+2t\mathbf{x})\hat{\sigma}_y\\&+&\frac{\left(\left(-1+2
e^{-t (\mathbf{y}+\mathbf{z})}\right) \mathbf{y}+\mathbf{z}\right)
\cos v}{\mathbf{y}+\mathbf{z}}\hat{\sigma}_z]\,.
\end{array}
\end{equation}

Expressions eq.\eqref{ADDensOperBlochthetaOmegatau} and
eq.\eqref{ADDensOperBlochxyz} are solutions of the  master
equation eq.\eqref{ADmastereq}. Unitary-equivalent Kraus matrices
eqs.\eqref{AD1}-\eqref{AD4} and eq.\eqref{GADKraus} describe the
same process.

For completeness, with the use of eqs.\eqref{ADEffects}, below we
compare the temporal behaviors  of the Bloch sphere for the
standard AD and  GAD channels.  Also we study the GAD channel
for various (high) temperatures via investigating temporal
behavior of the Bloch sphere volume.

\begin{figure}[!ht]
\centering
  \includegraphics[width=0.3\textwidth]{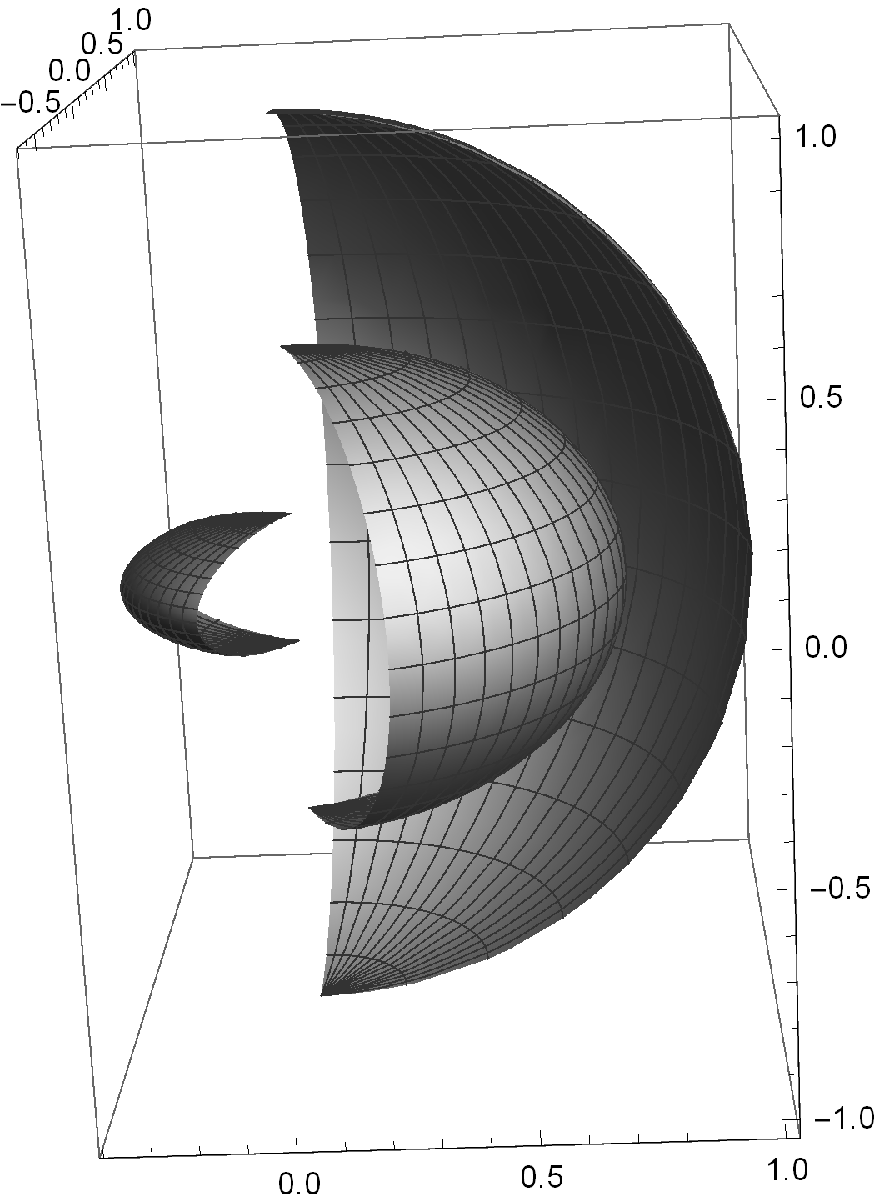}
  \includegraphics[width=0.3\textwidth]{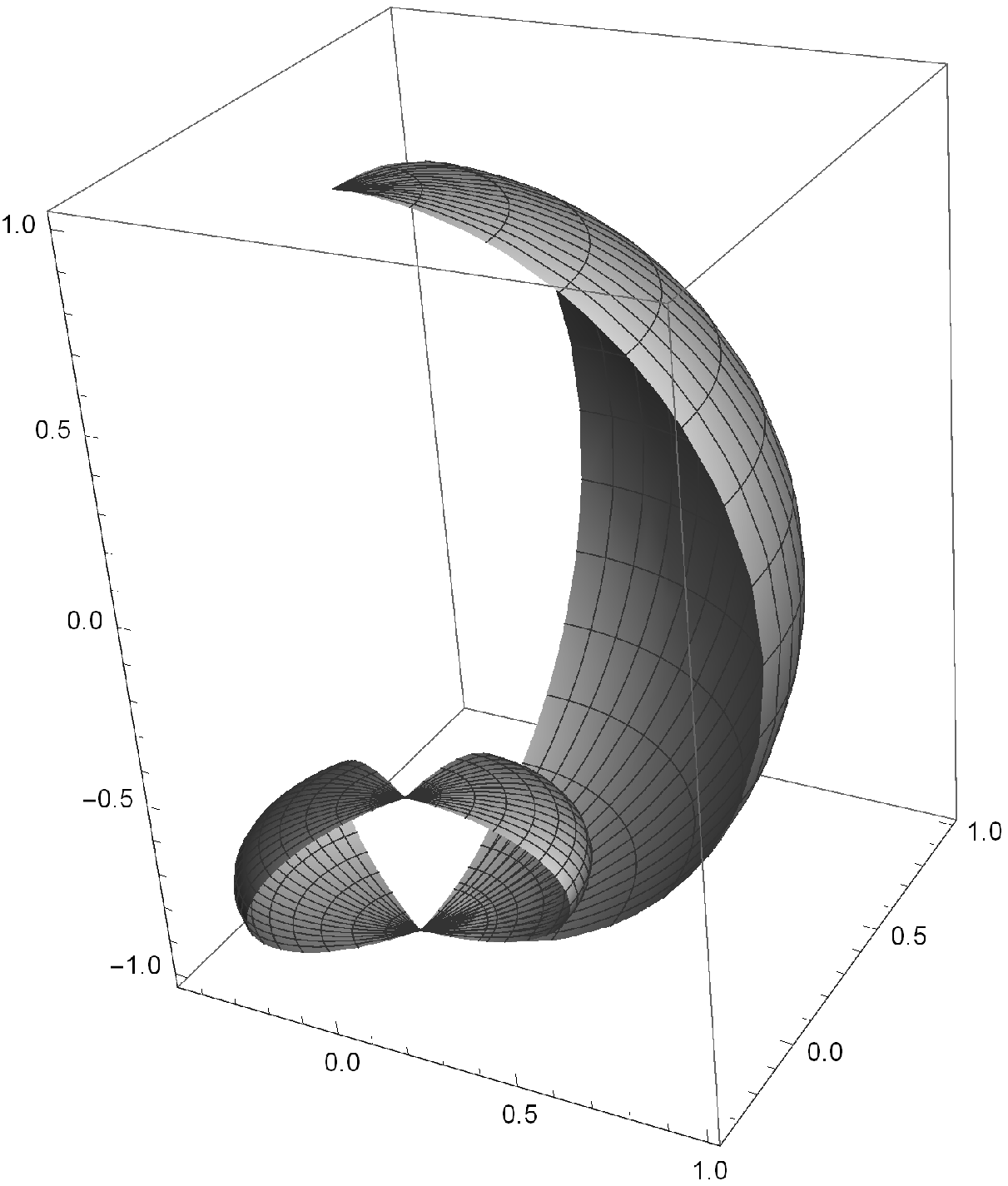}
\caption{(Left) Representation of the  GAD process  for high
temperature: the big sphere is for the initial instant of time
$t=0$--the unchanged Bloch sphere. For $t=0.05$, both smaller and larger ellipsoids  pertain to the GAD channel for $T=300$ and
$T=100$, respectively. (Right) A representation  for low
temperature ($T=1$):  the big sphere is for the initial instant of
time $t=0$. For $t=2.5$, there is only one ellipsoid for  both,
the standard AD and  the GAD channel, exhibiting their match in
this temperature regime. The  parameters: $\alpha=0.02$,
$\omega_0=10$ and $\omega_c=15$.}
  \label{FigAD3D}
\end{figure}

Fig.\ref{FigAD3D}(right) exhibits unchangeability of the ``ground'' state
$\vert 1\rangle$ on $T=0$K while Fig.1(left)  shows instability
(finite probability for excitation) of the ground state for the
finite temperature range; of course, this physical observation is
well known from the application of the quantum-optical master
equation to a two-state atom \cite{BreuPetr}. On the other hand,
Fig.\ref{FigAD3D}(left) reveals a faster change of the ``excited''
state $\vert 0\rangle$ for higher temperatures.

Time dependence of the Bloch-sphere volume is:
\begin{equation}\label{ADBlochvolumechange}
   V(\tau)=\frac{4\pi}{3}e^{-4\tau},
\end{equation}
equivalently $V(t)=\frac{4\pi}{3}e^{-2 (\mathbf{z}+\mathbf{y})t}$.
The relative change of the Bloch sphere volume,
$\kappa(t)=\frac{1}{V_0} \frac{dV(t)}{dt}$:
\begin{equation}\label{ADBlochspeedchange}
   \kappa(t)=-2(\mathbf{z}+\mathbf{y})e^{-2(\mathbf{z}+\mathbf{y})t}.
\end{equation}

\begin{figure}[!ht]
\centering
\includegraphics[width=0.4\textwidth]{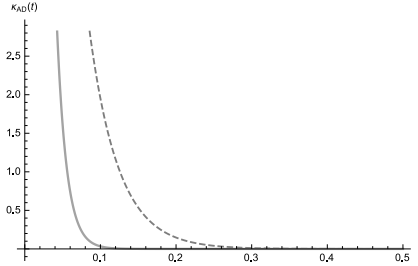}%
\caption{The relative change of the Bloch sphere volume for the
high-temperature GAD process. The dashed line is for  $T=100$  and
the thick one for  $T=300$. The parameters:
$\alpha=0.02$, $\omega_0=10$ and $\omega_c=15$.} \label{FigVADRel}
\end{figure}

Eq.\eqref{ADBlochspeedchange} regarding GAD process is presented
in Fig.\ref{FigVADRel};

Finally, GAD Kraus operators, eqs.\eqref{AD1}-\eqref{AD4} and
eq.\eqref{GADKraus}, can be shown to take the following forms in
the asymptotic limit , $\tau\to\infty$:

\begin{equation}\label{ADasym1}
  \mathbb{E}_1^{\text{Asym}}=\left(
\begin{array}{cc}
 0 & 0 \\
 \frac{i \sqrt{2-\Omega }}{2}  & 0 \\
\end{array}
\right)
\end{equation}

\begin{equation}\label{ADasym2}
  \mathbb{E}_2^{\text{Asym}}=\left(
\begin{array}{cc}
 0 & -\frac{i \sqrt{\Omega +2}}{2}  \\
 0 & 0 \\
\end{array}
\right)
\end{equation}

\begin{equation}\label{ADasym3}
   \mathbb{E}_3^{\text{Asym}}=\left(
\begin{array}{cc}
 0 & 0 \\
 0 & -\frac{\sqrt{2-\Omega }}{2} \\
\end{array}
\right)
\end{equation}

\begin{equation}\label{ADasym4}
  \mathbb{E}_4^{\text{Asym}}=\left(
\begin{array}{cc}
 \frac{\sqrt{\Omega +2}}{2} & 0 \\
 0 & 0 \\
\end{array}
\right)
\end{equation}
from which it is clear that, asymptotically, the actions of the
standard AD channel, eqs.(\eqref{AD1Stand}-\eqref{AD4Stand}), and
of the GAD channel,  eqs.(\eqref{ADasym1}-\eqref{ADasym4}) , are
not mutually equivalent, except for the choice $\Omega = -2$ for
the GAD channel.

\vspace{30mm}

\centerline{\bf 4. THE PHASE DAMPING  CHANNEL}
\vspace{12mm}

The phase damping (PD) quantum channel  models pure decoherence
without loss of energy for a single-qubit system. The Hamiltonian
for the total (closed) system is given by \cite{RivasHuelga}:
\begin{equation}\label{PDHamiltonian}
\hat{H}=\frac{\omega_0}{2}\hat{\sigma}_z+\int^{\omega_{max}}_0
d\omega \hat{a}^{\dag}_{\omega}\hat{a}_{\omega}+
\hat{\sigma}_z\otimes\int^{\omega_{max}}_0 d\omega h(\omega)
(\hat{a}^{\dag}_{\omega}+\hat{a}_{\omega}).
\end{equation}
Notation and the meaning of the terms in eq.\eqref{PDHamiltonian}
are the same as in eq.\eqref{ADHamiltonian}.

The  model eq.\eqref{PDHamiltonian} gives rise to the following
microscopic Markovian master equation in the interaction picture
\cite{RivasHuelga}:
\begin{equation}\label{PDmastereq}
\frac{d\hat{\rho}_S(t)}{dt}=r(
   \hat{\sigma}_z\hat{\rho}_S(t)\hat{\sigma}_z-\hat{\rho}_S(t)),
   \end{equation}
whereby the decay rate $r$  \cite{RivasHuelga}:
\begin{equation}\label{PDDampRate}
r=2\pi\lim_{\omega\to 0} J(|\omega|)\langle n(|\omega|)\rangle,
   \end{equation}
under assumption $\lim_{\omega\to 0} J(|\omega|)=0$. $J(\omega)$
is the spectral density of the bath  while $ \langle
n(\omega)\rangle$ is the mean number of the bosons for the thermal
state of the bath with the frequency $\omega$.

Following the recipe of    Section 2:
\begin{equation}\label{PDmatL}
L=\left(\begin{array}{cccc}
  0 & 0 & 0 & 0 \\
  0 & -2r & 0 & 0 \\
  0 & 0 & -2r & 0 \\
  0 & 0 & 0 &0 \\
\end{array}\right),
\end{equation}

\begin{equation}\label{PDmatF}
F=\left(\begin{array}{cccc}
  1 & 0 & 0 & 0 \\
  0 & e^{-2rt} & 0 & 0 \\
  0 & 0 & e^{-2rt} & 0 \\
  0 & 0 & 0 &1 \\
\end{array}\right),
\end{equation}

\begin{equation}\label{PDmatS}
S=\left(
\begin{array}{cccc}
 1+e^{-2 r t} & 0 & 0 & 0 \\
 0 & 0 & 0 & 0 \\
 0 & 0 & 0 & 0 \\
 0 & 0 & 0 & 1-e^{-2 r t}
\end{array}
\right),
\end{equation}
which give rise to the following Kraus operators:
\begin{equation}\label{PD1}
\mathbb{E}_1=\left(
\begin{array}{cc}
 \frac{\sqrt{1-e^{-2 r t}}}{\sqrt{2}} & 0 \\
 0 & -\frac{\sqrt{1-e^{-2 r t}}}{\sqrt{2}}
\end{array}
\right),
\end{equation}

\begin{equation}\label{PD2}
\mathbb{E}_2=\left(
\begin{array}{cc}
 \frac{\sqrt{1+e^{-2 r t}}}{\sqrt{2}} & 0 \\
 0 & \frac{\sqrt{1+e^{-2 r t}}}{\sqrt{2}}
\end{array}
\right).
\end{equation}

These matrices are the
$\hat{\sigma}_z=|0\rangle\langle0|-|1\rangle\langle1|$
representations of the well known Kraus operators for the PD
channel \cite{RomLoFranco}:
\begin{equation}\label{StandardPD}
\hat{E}_{0}=\sqrt{1-\frac{p(t)}{2}} \hat{I}, \hspace{12pt}
\hat{E}_{1}=\sqrt{\frac{p(t)}{2}}\hat{\sigma}_z
\end{equation}
where $p(t)\equiv1-e^{-2rt}$  while the completeness relation $
\sum_{k}\hat{E}_k(t)^{\dag}\hat{E}_k(t)=\hat{I}$ is satisfied.

From Kraus operators, eqs.\eqref{PD1} and \eqref{PD2} easily follows
\begin{subequations}
\label{PDEffects}
\begin{align}
\phi_{\tau}(\hat{I})&=\hat{I}\,, \label{PDEffectsA} \\
\phi_{\tau}(\hat{\sigma}_x)&=e^{-2rt}\hat{\sigma}_x\,, \label{PDEffectsB}\\
\phi_{\tau}(\hat{\sigma}_y)&=e^{-2rt}\hat{\sigma}_y\,, \label{PDEffectsC}\\
\phi_{\tau}(\hat{\sigma}_z)&=\hat{\sigma}_z\,.
\label{PDEffectsD}
\end{align}
\end{subequations}

Hence the solution of eq.\eqref{PDmastereq}:
\begin{equation}\label{PDDensOperBlochTR}
\begin{array}{r c l}
  \phi_{\tau}(\hat{\rho})&=&\frac{1}{2}[\hat{I}+e^{-2rt}\sin v\cos u\hat{\sigma}_x
  +e^{-2rt}\sin v\sin u\hat{\sigma}_y\\&+&\cos
  v\hat{\sigma}_z]\,.
  \end{array}
\end{equation}
for the initial state $\hat{\rho}=\frac{1}{2}\left(\hat{I}
+\vec{n} \cdot \hat{\vec{\sigma}}\right)$; $n = (n_x, n_y, n_z)$. Notice diagonalizability
of the state eq.\eqref{PDDensOperBlochTR} for long times ($t\rightarrow\infty$) in the $\hat{\sigma}_z$ eigenbasis, which
becomes the ``pointer basis'' for the decoherence process \cite{N&Ch} induced by the environment.
\vspace{20mm}

\centerline{\bf 5. CONCLUSION}
\vspace{12mm}

Detailed microscopic analysis of the differential form of the
amplitude damping and phase damping processes on a single
qubit gives rise to the Kraus operators that describe exactly the same process as the standard Kraus operators widely used for these processess.

\end{document}